\documentclass[aps,pre,preprint,amsmath,amssymb,superscriptaddress,showpacs]{revtex4-1}
\usepackage{amssymb}
\usepackage{color}
\usepackage{graphicx}
\usepackage{amsmath}
\usepackage{mathrsfs}
\usepackage{times}
\usepackage{subeqnarray}
\usepackage{cases}
\usepackage{bm}
\usepackage{float}

\begin{document}

\title{Cluster synchronization in complex network of coupled chaotic circuits:\\ an experimental study}

\author{Ben Cao}
\affiliation{School of Physics and Information Technology, Shaanxi Normal University, Xi'an 710062, China}
\affiliation{School of Physical Education, Shaanxi Normal University, Xi'an 710062, China}

\author{Yafeng Wang}
\affiliation{School of Physics and Information Technology, Shaanxi Normal University, Xi'an 710062, China}

\author{Liang Wang}
\affiliation{School of Physics and Information Technology, Shaanxi Normal University, Xi'an 710062, China}

\author{Yizhen Yu}
\affiliation{School of Physics and Information Technology, Shaanxi Normal University, Xi'an 710062, China}

\author{Xingang Wang}
\email[Email address: ]{wangxg@snnu.edu.cn}
\affiliation{School of Physics and Information Technology, Shaanxi Normal University, Xi'an 710062, China}

\begin{abstract}
By a small-size complex network of coupled chaotic Hindmarsh-Rose circuits, we study experimentally the stability of network synchronization to the removal of shortcut links. It is shown that the removal of a single shortcut link may destroy either completely or partially the network synchronization. Interestingly, when the network is partially desynchronized, it is found that the oscillators can be organized into different groups, with oscillators within each group being highly synchronized but are not for oscillators from different groups, showing the intriguing phenomenon of cluster synchronization. The experimental results are analyzed by the method of eigenvalue analysis, which implies that the formation of cluster synchronization is crucially dependent on the network symmetries. Our study demonstrates the observability of cluster synchronization in realistic systems, and indicates the feasibility of controlling network synchronization by adjusting network topology. 
\end{abstract}
\date{\today}
\maketitle

\section{Introduction}
The collective dynamics of coupled nonlinear oscillators has received continuous interest by researchers from different fields \cite{Kuramoto,PRK,Strogatz,Boccaletti2006,SYNREVArenas}. For systems of coupled identical chaotic oscillators, it has been shown that, despite of the intrinsic sensitivity of the system dynamics, the trajectories of the coupled oscillators can be converged to a single one when the coupling strength is larger than some threshold, namely the state of complete synchronization \cite{CS1990}. Complete synchronization also emerges when an ensemble of chaotic oscillators are coupled on regular structures, e.g., the global (all-to-all) coupling structure or the lattices \cite{Boccaletti2006}. For diffusive couplings, the critical coupling for synchronization could be analyzed by the method of master stability function (MSF)
 \cite{MSF1,MSF2,MSF3}, which decouples the system into isolated modes and shows that the system's synchronizability is jointly affected by the MSF curve (determined by the local dynamics and the coupling function) and the eigenvalues of the coupling matrix (determined by the coupling structure). Whereas most studies of chaos synchronization focus on the critical point where all oscillators are synchronized to the same trajectory, i.e., the global synchronization state, there are also studies on the collective behaviors of the coupled oscillators before this critical point, i.e., the transition to global synchronization \cite{Boccaletti2006}. In exploring the transition process of synchronization, an interesting phenomenon observed commonly in different types of systems is that the oscillators could be self-organized into different synchronous clusters, namely the cluster synchronization (also known as group or partial synchronization) \cite{CSHansel1993,CSBelykh,CSHasler,CSYZHANG,CSPikovsky,CSExpZY,CSBAO,CSCSZ,CSJZ}. In cluster synchronization, oscillators within the same clusters are highly or even completely synchronized, but are not for oscillators from different clusters. For the regular coupling structures, cluster synchronization is normally generated by the mechanism of symmetry breaking \cite{Golubitsky1985}, and the spatial distribution of the clusters, i.e., the synchronous pattern, in general can be analyzed in the mode space by the MSF method \cite{SynPatPecora1998}. Since cluster synchronization is often regarded as the intermediate states between non-synchronous and globally synchronous states, it thus underpins the bifurcation diagrams of synchronization transition \cite{Boccaletti2006,SYNREVArenas}. Besides theoretical interest, cluster synchronization is also of great importance to the functionality of many biological and engineering systems in practice, e.g., the cognition of human brain relies on the synchronous firing of neural clusters distributed over the cortical cortex \cite{BOOKBrain}. 

The discovery of the small-world and scale-free features in many natural and man-made networks \cite{NETSW,NETSFN} have stirred a new surge of studies on the synchronization of complex networks in the past two decades \cite{NETSYNGade,NETSYNCKHU2000,NETSYNBarahona,NETSYNNishikawa2003,NETSYNMotter2005,NETSYNWXG2007,NETSYNCBF2012,NETSYNHZW,NETSYNWY2015,NETSYNMotter2016}. In the study of complex network synchronization, one of the central topics is the interplay between the network dynamics and structure \cite{Boccaletti2006,SYNREVArenas}. It has been shown that by introducing randomly a few of shortcut links onto a regular network (so as to generate the small-world feature), the network synchronizability can be significantly enhanced \cite{NETSYNGade,NETSYNCKHU2000,NETSYNBarahona} and, by introducing weights to the network links, the synchronizability of scale-free networks can be higher than that of small-world networks (of the same network size and connectivity) \cite{NETSYNNishikawa2003,NETSYNMotter2005,NETSYNWXG2007}. Similar to the studies on regular networks, studies on the synchronization of complex networks focus on still the stability of the global synchronization state, which, for the case of identical oscillators and diffusive couplings, can be analyzed by the MSF method. Therefore, from the viewpoint of the MSF analysis, the impact of network structure on synchronization seems to be reflecting in only the modified eigenvalues calculated from the coupling matrix. With this understanding, in the past years a number of studies had been conducted on the optimization of network synchronization \cite{Boccaletti2006}, with the approaches proposed including designing new network structures, adjusting the weighting schemes, and adopting different coupling strategies. At first glance, it seems that complex network and cluster synchronization are incompatible with each other, as random shortcuts will destroy synchronous clusters \cite{PATTERNPARK}. Yet recent studies give the accumulating evidences which show that cluster synchronization is also observable in complex networks \cite{CSBAO,CSCSZ,CSOTT2007,CSBelykh2,CSRusso,CSDahms,CSNicosia,PATTERNCONTROLWXG,CSWilliams2013,CSWXG2014,CSPecora2014,CSFrancesco,SynPatBansal,SynPatMotter,SynPatSchaub,SynPatPecora,CSMotter2017}. The first batch of evidences come from the synchronization of small-size complex network possessing reflection symmetries \cite{CSYZHANG,CSBAO,CSJZ,PATTERNCONTROLWXG,CSWXG2014}, where it is found, in spite of the presence of random shortcut links, the oscillators can be synchronized in pairs according to the network reflection symmetries. Additional evidences from large-size complex network possessing permutation symmetries have been also provided \cite{CSPecora2014,CSFrancesco,SynPatBansal,SynPatMotter,SynPatSchaub,SynPatPecora}. With the help of computational group theory algorithm \cite{CGT}, the permutation symmetries of large-size complex networks now can be identified numerically, which, combining with the generalized method of eigenvalue analysis \cite{CSFrancesco,SynPatMotter,LWJ1,LWJ2,CSMotter2017}, can be used to analyze the formation of cluster synchronization in the general complex networks.

Despite the theoretical evidences, cluster synchronization is hardly observed in experiment, for its sensitivity to the parameter mismatch among the oscillators and the noise perturbations \cite{NonIdenticalSJ2009,SynPatPecora}. So far, the widely cited experimental results on cluster synchronization are carried out on a small-size complex network of coupled optoelectronic oscillators \cite{CSPecora2014,CSFrancesco}, in which the nodal dynamics is described by discrete maps and the couplings are realized artificially with the help of computer software. As oscillators in realistic systems are mostly represented by time-continuous differential equations, it is still intriguing to see whether cluster synchronization can be observed in complex network consisting of time-continuous oscillators. Moreover, in the optoelectronic experiment the couplings of the maps are implemented through the computer software \cite{CSPecora2014,CSFrancesco}, which is also not a suitable description for the natural systems. With these concerns, we study in the present work the generation of cluster synchronization in complex network constituted by chaotic neural circuits, with the dynamics of the neural circuits described by time-continuous differential equations and the circuits are coupled through hardwares. Specifically, starting from a synchronizable complex network, we remove one of the shortcut links in the network, and investigate the responses of the system dynamics to this removal. Our main finding is that, depending on the specific link being removed, the network may either partially or globally desynchronized. In the case of partial desynchronization, remarkably, it is found that the neural oscillators are synchronized into different groups, i.e., showing the phenomenon of cluster synchronization. Our study shows that, despite the parameter mismatches and noise perturbations, cluster synchronization can still be generated in realistic systems of networked nonlinear oscillators \cite{SynPatPecora}. This finding paves a way to the study of more complicated synchronous clusters in realistic systems, e.g., the synchronous firing patterns in complex neural networks.

The rest of the paper is organized as follows. In Sec. II, we will present the model of networked chaotic circuits and describe in detail the hardware realization of the electrical neural circuits. In Sec. III, we will present the experimental results. In Sec. IV, we will first present the theoretical framework for analyzing the stability of cluster synchronization and derive explicitly the stability criteria, then use this theoretical framework to interpret the experimental results. Discussions and conclusion will be given in Sec. V.

\section{Model and experimental setup}

We consider the following general model of networked nonlinear oscillators \cite{NETSYNGade,NETSYNCKHU2000,NETSYNBarahona,NETSYNNishikawa2003,NETSYNMotter2005,NETSYNWXG2007,NETSYNMotter2016}, 
\begin{equation}
\dot{\mathbf{x}}_i=\mathbf{F}(\mathbf{x}_i)+\varepsilon\sum\limits^{N}_{j=1}a_{ij}[\mathbf{H}(\mathbf{x}_j)-\mathbf{H}(\mathbf{x}_i)],
\label{model}
\end{equation}
with $i,j=1,2,\ldots,N$ the oscillator (node) indices and $N$ the number of oscillators (the network size). $x_{i}$ describes the dynamical variable vector of the \emph{i}th oscillator, and $\varepsilon$ is the uniform coupling strength. The coupling relationship of the oscillators, i.e., the network structure, is characterized by the adjacent matrix $A=\{a_{ij}\}$, with its element defined as follows: $a_{ij}=a_{ji}=1$ if nodes $i$ and $j$ is connected by a link in the network, and $a_{ij}=0$ otherwise. The nodal dynamics in the isolated form is described by the function $\mathbf{F}$, which is identical for the oscillators in the theoretical studies but is not in the experimental realizations. $\mathbf{H}$ denotes the coupling function. Eq. (\ref{model}), or its equivalent forms, describes the dynamics of a large variety of spatiotemporal systems, and has been employed as one of the standard models in exploring the synchronization behaviors of coupled oscillators \cite{CSPecora2014}.

To be concrete, we adopt the Hindmarsh-Rose (HR) oscillator as the neuron dynamics \cite{HRmodel,HRcircuit}, which in the isolated form is described by the equations
\begin{eqnarray}
\begin{cases}
\dot{x}=y+\phi(x)-z+I,\\
\dot{y}=\psi(x)-y,   \\
\dot{z}=r[s(x-x_R)-z],\\
\end{cases}
\label{eq2}
\end{eqnarray}
with
\begin{equation}
\phi(x)=-ax^3+bx^2, \nonumber \\
\end{equation}
and
\begin{equation}
\psi(y)=c-dx^2.  \nonumber 
\end{equation}
Here, $x$ represents the membrane potential, $y$ represents the transport rates of the fast ion channels (the spiking variable), and $z$ represents the transport rate of the slow ion channels (the bursting variable). $I$ describes the external stimulating current, which is used as the bifurcation parameter to adjust the neuron dynamics. In theoretical studies, we set the nodal parameters as $(a,b,c,d,r,s,x_R,I)=(1,3,1,5,6\times 10^{-3},4,-1.6,3.2)$, by which the isolated neuron shows the chaotic bursting behavior \cite{HRmodel}. The coupling function is chosen as $\bm{H}(\bm{x})=[x,0,0]^T$, i.e., neurons are coupled through their membrane potentials (i.e., the gap junctions).

\begin{figure*}[tbp]
\includegraphics[width=\textwidth]{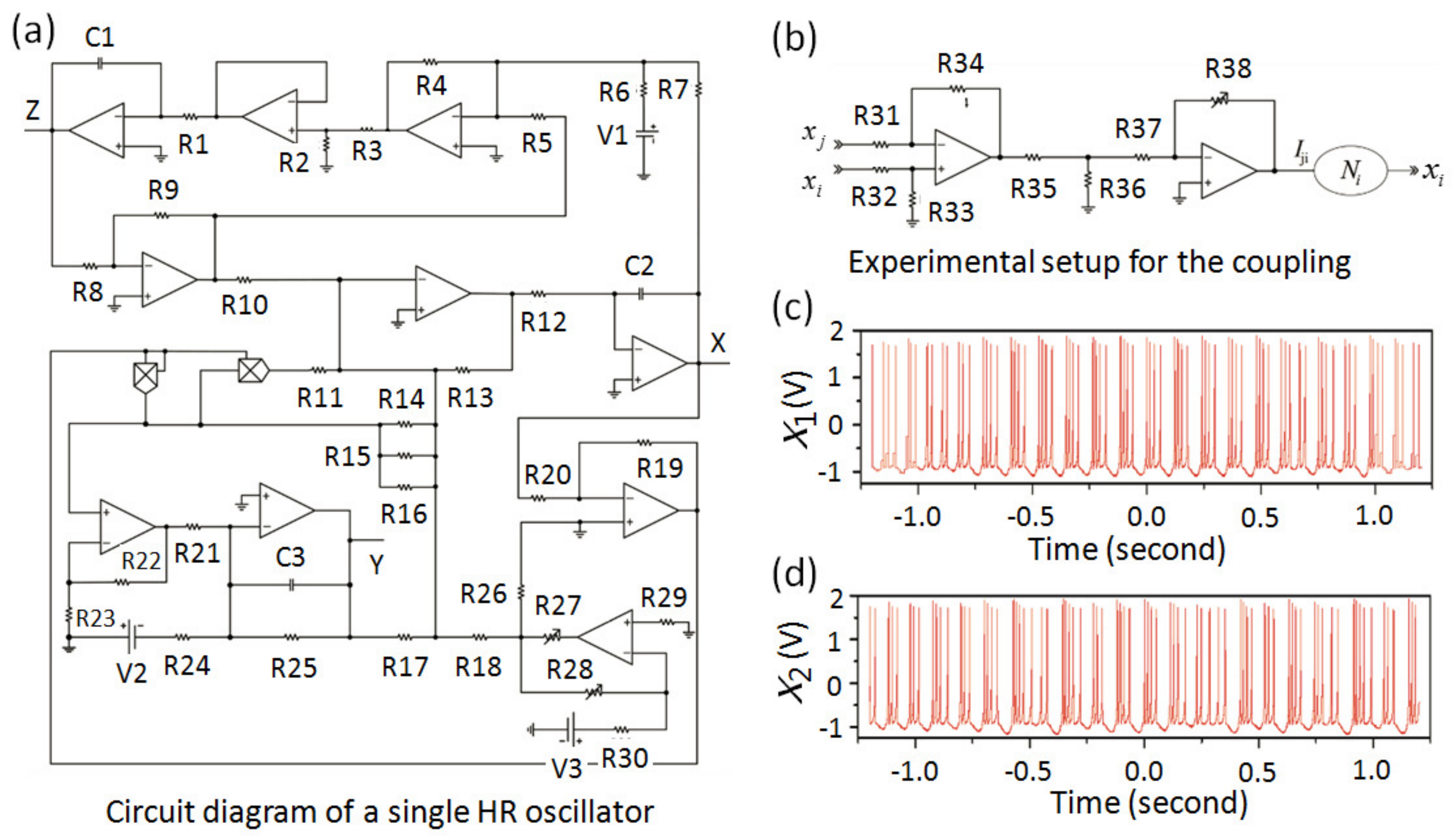}
\caption{(Color online) Experimental setup for the HR circuit. (a) Schematic diagram of a single HR circuit. AD633 and TL084CN denote, respectively, the analog multipliers and the operational amplifiers, which are fed with a $\pm12V$ power supply. The voltages of the direct-current sources are $(V_{1},V_{2},V_{3})=(1.6\ V,1.5\ V,12\ V)$. By tuning the parameters $(R_{27},R_{28})$, the circuit could present the quiescent, periodic spiking, and chaotic bursting behaviors. (b) Schematic diagram of the coupling circuit. The coupling strength is controlled by tuning $R_{38}$. (c-d) By the parameters $(R_{27},R_{28})=(1.5\ k\Omega,0.5\ k\Omega)$, the time evolution of the $x$-component (voltage) of two isolated HR circuits recorded by a two-channel digital oscilloscope (Rigol DS1052E, $50$\ MHz, 1GS/s), which show the typical chaotic bursting behaviors. See the context for the detailed description of the circuits.}
\label{fig1}
\end{figure*}

In experiment, the HR oscillator is modeled by the electrical circuit schematically plotted in Fig. \ref{fig1}(a). The dynamics of the circuit is described by the equations
\begin{eqnarray}
\begin{cases}
C_{2}R_{12}\dot{x} =\frac{R_{13}}{R_{17}}y-\frac{R_{13}}{R_{11}}x^3+\frac{R_{13}}{R_{14}\shortparallel R_{15}\shortparallel R_{16}}x^2-\frac{R_{10}R_{13}}{R_{8}R_{9}}z+I,\\
C_{3}R_{25}\dot{y} =\frac{R_{2}}{R_{3}}(1+\frac{R_{22}}{R_{23}})x^2-\frac{R_{25}}{R_{24}}V_{2}-\frac{R_{25}}{R_{21}}y,\\
C_{1}R_{1}\dot{z} =\frac{R_{2}}{R_{3}+R_{2}}(\frac{R_{4}}{R_{7}}x+\frac{R_{4}}{R_{6}}V_{1}-\frac{R_{9}}{R_{8}}z),\\
\end{cases}
\end{eqnarray}
with
\begin{equation}
I =\frac{1}{R_{30}}(1+\frac{R_{28}}{R_{26}})V_{3}.\nonumber \\
\end{equation}
In the circuit, all the operations (addition, subtraction, multiplication and integration) are realized by amplifiers, and, by tuning the resistors $R_{27}$ and $R_{28}$, the circuit can present rich dynamics. In our study, the parameters of the circuit elements are chosen as $C_1=C_2=C_3=47\ nF$, $R_{2}=R_{17}=R_{26}=R_{29}=R_{30}=1\ k\Omega$, $R_{6}=R_{7}=R_{23}=2.5\ k\Omega$, and $R_{4}=R_{5}=R_{8}=R_{9}=R_{10}=R_{11}=R_{13}=R_{14}=R_{15}=R_{16}=R_{17}=R_{18}=R_{22}=10\ k\Omega$. Setting $(R_{27},R_{28})=(1.5\ k\Omega,0.5\ k\Omega)$, the circuit generates robust chaotic bursting behaviors, as shown in Figs. \ref{fig1}(c-d) for two of the five circuits assembled in our lab. It is worth noting that the circuit elements are of $5\%$ parameter errors, which makes the assembled circuits essentially non-identical. Despite the parameter errors, chaotic bursting is robustly generated for all five circuits, which has been checked individually before coupling them into a network. The couplings between the oscillators are realized by the circuit plotted in Fig. \ref{fig1}(b), which is constituted by the subtraction circuit, the attenuating circuit, and the reverse amplification circuit. This circuit couples the $x$-components (voltages) of the oscillators in a linear fashion, $\mathbf{H}=[x,0,0]^T$, with the coupling strength $\varepsilon^{exp}\propto R_{38}$. In experimental studies, we fix the parameters of the coupling circuits as $R_{31}=R_{32}=R_{33}=R_{34}=R_{36}=100\ k\Omega$, $R_{35}=500\ k\Omega$ and $R_{37}=10\ k\Omega$, while adjusting the coupling strength by changing $R_{38}$ (which gives the experimental coupling strength $\varepsilon^{exp}\approx 0.027\times R_{38}$). 

\section{Experimental results}

\begin{figure*}[tbp]
\includegraphics[width=\textwidth]{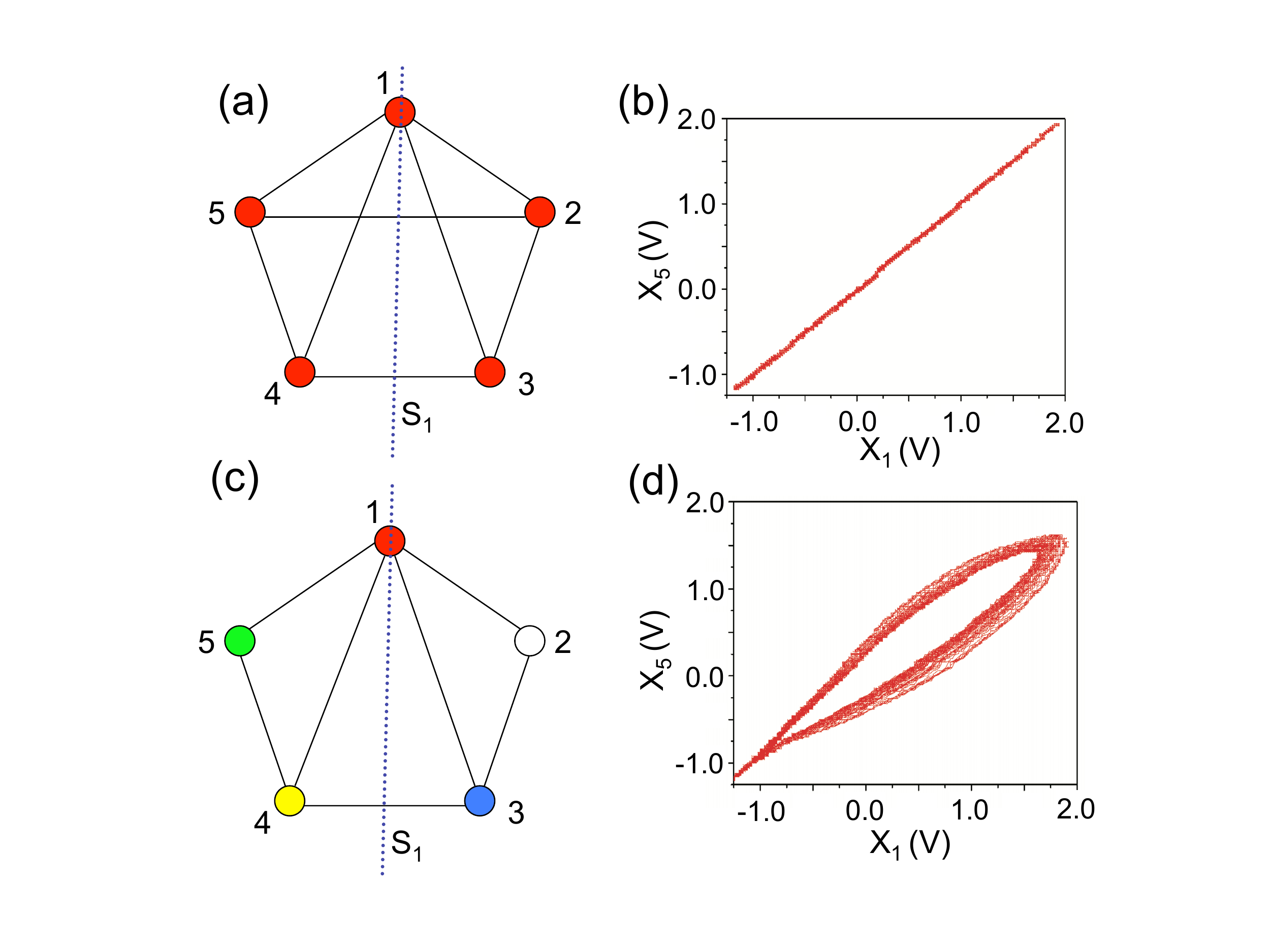}
\caption{(Color online) Experimental results for network synchronization. (a) The structure of the original network. The network reaches global synchronization when $R_{38}>12.1\ k\Omega$. (b) $x_5$ versus $x_1$ for $R_{38}=12.8\ k\Omega$ ($\varepsilon^{exp} \approx 0.33$). (c) The structure of the modified network, in which the link between nodes $2$ and $5$ is removed. (d) For $R_{38}=12.8\ k\Omega$ in the modified network, $x_5$ versus $x_1$. The network is completely desynchronized. $\mathbf{S}_1$ denotes the reflection symmetry satisfied by the network structures. Nodes of the same color are of the same trajectory. For the network in (a), we have $\lambda_{max}^B=\lambda_2^D=-3$ and $\varepsilon>\varepsilon_c\approx 0.59$, which, according to the eigenvalue analysis, lead to global synchronization. For the network in (c), we have $\lambda_{max}^B=-1.6$, which is larger than $\lambda_2^D=-3$. According to the eigenvalue analysis, the pattern $(a,b,c,c,b)$ associated to symmetry $\mathbf{S}_1$ is stable in the range $\varepsilon\in(0.31,0.59)$.}
\label{fig2}
\end{figure*}

We now couple the circuits into a network and investigate their collective behaviors experimentally. The network structure is shown in Fig. \ref{fig2}(a), which is constructed by adding randomly three shortcut links onto a ring of $5$ nodes. This network, although of small size, has the combined feature of regular and complex networks. In specific, although the shortcut links destroy the rotation symmetry (associated with the ring structure), the network possesses still the reflection symmetry (denoted by $\mathbf{S}_1$). As such, this network could be used as an intermediate structure between the regular and complex networks. In the meantime, because of the small network size, we are able to monitor the time evolutions for all oscillators simultaneously in experiment, and analyze the synchronization relationship among the oscillator timely. 

To search for cluster synchronization, we fix $R_{37}=10\ k\Omega$ while increasing $R_{38}$ from $0$ gradually, which corresponds to the increase of the uniform coupling strength $\varepsilon$ in the theoretical model described by Eq. (\ref{model}). Experimental data shows that as $R_{38}$ increases from $0$ to $12.1\ k\Omega$, no synchronization is established between any pair of oscillators in the network; and the network is globally synchronized when $R_{38}>12.1\ k\Omega$ ($\varepsilon^{exp}_c\approx 0.33$). By $R_{38}=12.8\ k\Omega$ ($\varepsilon^{exp}\approx 0.35$), we plot in Fig. \ref{fig2}(b) the $x$-components of two representative oscillators, $x_1(t)$ and $x_5(t)$, which shows that the two components are well synchronized. This synchronization behavior is observed for any pair of oscillators in the network, confirming that the network is globally synchronized under this coupling strength. 

We next try a different approach for generating cluster synchronization: adjusting the network structure. In doing so, we fix the coupling strength, $R_{38}=12.8\ k\Omega$, and remove one of the shortcut links in the network. This approach is motivated by the recent studies of network synchronization \cite{NETSYNGade,NETSYNCKHU2000,NETSYNBarahona,NETSYNNishikawa2003,NETSYNMotter2005,NETSYNWXG2007,NETSYNCBF2012,PATTERNCONTROLWXG,NETSYNHZW,NETSYNWY2015,NETSYNMotter2016}, which show that, in analogy to the conventional bifurcation parameters of nonlinear systems, network topology can be regarded as an alternative bifurcation parameter for tuning the network dynamics. In particular, the synchronizability of a complex network can be significantly affected by adding or removing a few of the shortcut links in a complex network. As by $R_{38}=12.8\ k\Omega$ the network is globally synchronizable, the removal of shortcut link thus is expected to deteriorate the global synchronization. Yet it remains not clear whether this approach is able to generate cluster synchronization. Removing the shortcut link between oscillators $2$ and $5$ [as shown in Fig. \ref{fig2}(c)], we plot in Fig. \ref{fig2}(d) $x_1(t)$ versus $x_5(t)$. It is seen that the two oscillators are desynchronized from each other. By checking the synchronization relationship among all the oscillators, it is further revealed that no synchronization is established between any pair of the oscillators, indicating that the network is completely desynchronized by removing the shortcut link between oscillators $2$ and $5$.   

\begin{figure*}[bpt]
\includegraphics[width=\textwidth]{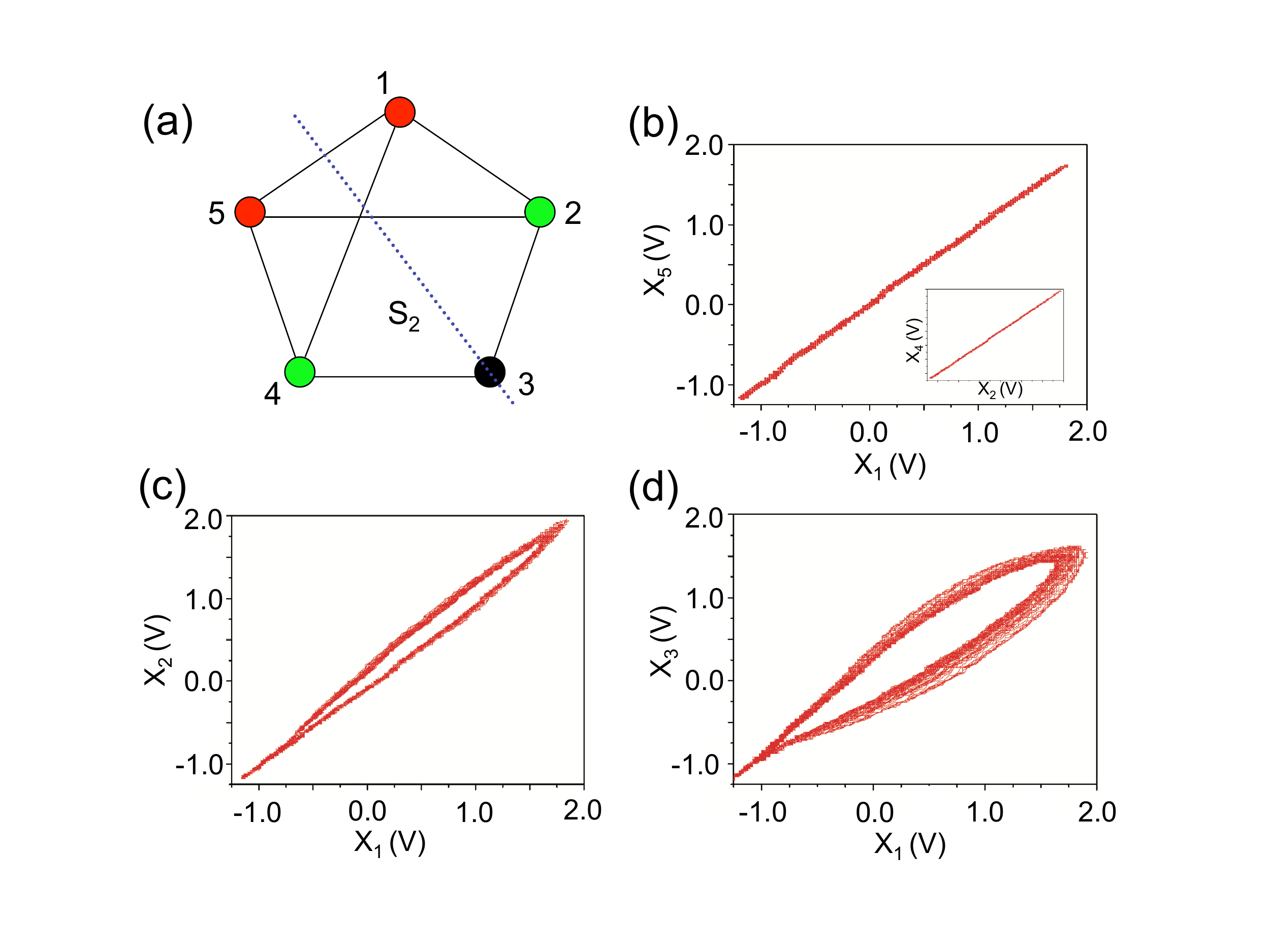}
\caption{(Color online) Experimental observation of cluster synchronization when the shortcut link between nodes $1$ and $3$ is removed. (a) The structure of the modified network, which satisfies the reflection symmetry $\mathbf{S}_2$. (b-d) Representative synchronization relationships among the oscillators. (b) $x_5$(t) versus $x_1$(t). Inset: $x_4$(t) versus $x_2$(t). (c) $x_2$(t) versus $x_1$(t). (d) $x_3$(t) versus $x_1$(t). Synchronization is observed between oscillators $1$ and $5$, and between oscillators $2$ and $4$, i.e., the system reaches the synchronous pattern $(a,b,c,b,a)$. Nodes of the same color belong to the same synchronous cluster.}
\label{fig3}
\end{figure*}

We move on to generate cluster synchronization by trying other shortcut links. As the removal of the link between nodes $1$ and $3$ results in the same network structure as the removal of the link between nodes $1$ and $4$, we therefore only need to investigate one case. Removing the link between nodes $1$ and $3$ [Fig. \ref{fig3}(a) shows the structure of the modified network, which satisfies the reflection symmetry $\mathbf{S}_2$], we plot in Fig. \ref{fig3}(b) again $x_5(t)$ versus $x_1(t)$. It is seen that the synchronization between the two oscillators is maintained. By checking the synchronization relationship among all the oscillators, it is revealed that besides the pair of oscillators $(1,5)$, the oscillators $2$ and $4$ are also synchronized. However, the two synchronization pairs, $(1,5)$ and $(2,4)$, are of different trajectories, as depicted in Fig. \ref{fig3}(c). Furthermore, it is found that oscillator $3$ is not synchronized to any pair [see Fig. \ref{fig3}(d)]. We repeat the experiment for several times, and the same synchronization relationship always appears. We thus infer from the experimental observations that cluster synchronization is generated in the network of Fig. \ref{fig3}(a). 

\section{Theoretical analysis}

\subsection{Stability of cluster synchronization}\label{theory}

Why the removal of the shortcut link between nodes $2$ and $5$ results in complete desynchronization [Figs. \ref{fig2}(c-d)], while the removal of the shortcut link between nodes $1$ and $3$ leads to cluster synchronization? To find out the reasons, we proceed to conduct a theoretical analysis on the different roles of the shortcut links in generating cluster synchronization, based on the method of eigenvalue analysis \cite{CSWXG2014,SynPatMotter}. Firstly, it should be noted that the cluster synchronization states possibly generated in a network are closely related to the network topological symmetries. Specifically, for any given network symmetry, one can always generate the corresponding cluster synchronization state by setting the initial conditions of the symmetric nodes as identical (whatever the coupling strength). The cluster synchronization states so generated, however, might be unstable under noise perturbations. For realistic systems such as coupled electrical circuits, only stable cluster synchronization states are observable. The study of cluster synchronization thus is divided into two separating issues: (1) finding all the network symmetries and (2) analyzing the stability of each cluster synchronization state associated with the network symmetry. For the simple network plotted in Fig. \ref{fig1}(a), the network symmetry can be inspected straightforwardly; whereas for large-size complex networks, this can be done numerically by the approach of computational group theory \cite{CGT,CSPecora2014}. 

Grouping nodes according to the network symmetry (symmetric nodes are grouped into the same cluster), we are able to construct the corresponding cluster synchronization state. The stability of the cluster synchronization state can be analyzed, as follows. Assume that the network nodes are grouped into $M$ clusters and the size of the $m$th cluster is $C_m$, the evolutional equation of the $i$th oscillator belonging to the $m$th cluster can be written as
\begin{equation}
\dot{\mathbf{x}}_i=\mathbf{F}(\mathbf{x}_i)+\varepsilon\sum\limits_{j\in V_m}a_{ij}[\mathbf{H}(\mathbf{x}_j)-\mathbf{H}(\mathbf{x}_i)]+\varepsilon\sum\limits_{j\notin V_m}a_{ij}[\mathbf{H}(\mathbf{x}_j)-\mathbf{H}(\mathbf{x}_i)],
\label{eq4}
\end{equation}
with $V_m$ the set of nodes in cluster $m$. In the right-hand-side of Eq. (\ref{eq4}), the $2$nd term corresponds to couplings that node $i$ received from nodes within the same cluster, and the $3$rd term corresponds to couplings receives from nodes in other clusters. Let the network being initially staying at the cluster synchronization state and denote $\mathscr{X}_m$ as the synchronous manifold of the $m$th cluster, then Eq. (\ref{eq4}) can be rewritten as
\begin{equation}
\dot{\mathbf{x}}_i=\mathbf{F}(\mathscr{X}_m)+\varepsilon\sum\limits_{j\in V_m}a_{ij}[\mathbf{H}(\mathbf{x}_j)-\mathbf{H}(\mathbf{x}_i)]+\varepsilon\sum_{m'\neq m}\sum_{l\in V_{m'}}a_{il}[\mathbf{H}(\mathscr{X}_{m'})-\mathbf{H}(\mathscr{X}_m)].
\label{eq5}
\end{equation}
As $\mathbf{x}_i = \mathscr{X}_m$ for $i\in V_m$, the above equation can be simplified to
\begin{equation}
\dot{\mathscr{X}}_m=\mathbf{F}(\mathscr{X}_m)+\varepsilon\sum^M_{m'=1}a'_{mm'}[\mathbf{H}(\mathscr{X}_{m'})-\mathbf{H}(\mathscr{X}_m)],
\label{eq6}
\end{equation}
with $a'_{mm'}=\sum_{l\in V_{m'}}a_{il}$ the integrated coupling strength that node $i$ in cluster $m$ receives from all nodes in cluster $m'$. Eq. (\ref{eq6}) describes the interactions between the synchronous manifolds $\mathscr{X}_m$ (with $\mathscr{X}_m$ the unified motion of all oscillators in cluster $m$), and defines the dynamics of the cluster synchronization state in the absence of noise perturbations. Let $\delta \mathbf{x}_i=\mathbf{x}_i-\mathscr{X}_m$ be the infinitesimal perturbations added on oscillator $i$, then whether the cluster synchronization state described by Eq. (\ref{eq6}) is stable is determined by the following set of variational equations [obtained by linearizing Eq. (\ref{eq5}) around the cluster synchronous manifolds $\mathscr{X}_m$]
\begin{eqnarray}
\delta \dot{\mathbf{x}}_i=&&D\mathbf{F}(\mathscr{X}_m)\delta \mathbf{x}_i
+ \varepsilon \sum_{j\in V_m} a_{ij}D\mathbf{H}(\mathscr{X}_m)(\delta \mathbf{x}_j-\delta \mathbf{x}_i) \nonumber \\
&&+\varepsilon\sum_{m'\neq m}\sum_{l\in V_{m'}}a_{il}\left[ D\mathbf{H}(\mathscr{X}_{m'})\delta \mathbf{x}_{l}-D\mathbf{H}(\mathscr{X}_m)\delta \mathbf{x}_i\right],
\label{eq7}
\end{eqnarray}
with $D\mathbf{F}$ and $D\mathbf{H}$ the Jacobin matrices. For the cluster synchronization state to be stable, the necessary condition is that $\delta \mathbf{x}$ damps to $0$ with time for all the oscillators. 

Let $\mathbf{R}_{S}$ be the permutation matrix associated to the network symmetry $\mathbf{S}$, we can decouple the variational equations [Eq. (\ref{eq7})] by transforming them into the mode space spanned by the eigenvectors of $\mathbf{R}_{S}$. The permutation matrix can be constructed from the network symmetry, as follows. If nodes $i$ and $j$ are symmetric and belong to the same cluster, we set  $r_{ij}=r_{ji}=1$, otherwise $r_{ij}=r_{ji}=0$; if node $i$ does not belong to any of the clusters, i.e., it is defined as an isolated node, we set $r_{ii}=1$ and $r_{ij}=0$ for $i\neq j$. Denote $\mathbf{T}_s$ as the transformation matrix constructed from the eigenvectors of $\mathbf{R}_{S}$, we apply it to the coupling matrix $\mathbf{A}$ and will obtain the following blocked matrix
\begin{equation}
\mathbf{A}'=\mathbf{T}_s^{-1}\mathbf{A}\mathbf{T}_s=\left(
  \begin{array}{ccc}
    \mathbf{B} &      0     \\
         0     &  \mathbf{D} \\
  \end{array}
  \right).
 \label{eq8}
\end{equation}
The submatrix $\mathbf{D}$ has the dimensions $M\times M$, which characterizes the space of the synchronous manifolds [i.e., the dynamics described by Eq. (\ref{eq6})]. The submatrix $\mathbf{B}$ has the dimensions $(N-M)\times (N-M)$, which characterizes the space transverse to the synchronous manifolds. (Besides dimensions, the submatrix $\mathbf{D}$ is also distinguished from $\mathbf{B}$ by owing the null eigenvalue, as will be discussed later.) For modes in the synchronous space, we have the variational equations       
\begin{equation}
\delta \dot{\mathbf{y}}_{m}=D\mathbf{F}(\mathscr{X}_m)\delta \mathbf{y}_{m}+\varepsilon\sum_{m'=1}^M d_{mm'} [D\mathbf{H}(\mathscr{X}_{m'}) \delta \mathbf{y}_{m'} - D\mathbf{H}(\mathscr{X}_{m}) \delta \mathbf{y}_{m}],
\label{eq9a}
\end{equation}
with $m,m'=1,\ldots,M$. The transverse space is constituted by $M$ subspaces. For modes in the transverse space of the $m$th cluster, we have the variational equations  
\begin{equation}
\delta \dot{\mathbf{y}}^m_{i'}=D\mathbf{F}(\mathscr{X}_m)\delta \mathbf{y}_{m}+\varepsilon\sum_{j'=1}^{C_m-1} b^m_{i'j'} [D\mathbf{H}(\mathscr{X}_{m}) \delta \mathbf{y}_{j'}- D\mathbf{H}(\mathscr{X}_{m}) \delta \mathbf{y}_{i'}],
\label{eq9b}
\end{equation}
where $\mathbf{B}^m=\{b_{i'j'}\}$ characterizes the $(C_m-1)$-dimentional mode space transverse to $\mathscr{X}_m$. (Please note that the transverse space is constituted by all $M$ transverse subspaces, $\mathbf{B}=\sum_{m}\oplus\mathbf{B}^m$.) In Eqs. (\ref{eq9a}) and (\ref{eq9b}), $\Delta \mathbf{Y}=(\delta \mathbf{y}_1,\ldots,\delta \mathbf{y}_M, \delta \mathbf{y}_{M+1},\ldots,\delta \mathbf{y}_N)^T$ are the perturbation modes, which is obtained from the perturbation vector $\Delta\mathbf{X}=(\delta x_1,\ldots,\delta x_N)^T$ through the transformation $\Delta\mathbf{Y}=\mathbf{T}_s^{-1}\Delta\mathbf{X}$. To make the cluster synchronization state stable, the necessary condition now becomes that all the transverse modes of any cluster should be damping with time, i.e., the largest Lyapunov exponent of the dynamics described by Eq. (\ref{eq9b}) should be negative. 

As the transverse modes of a cluster are decoupled from those of other clusters [see Eq. (\ref{eq9b})], the stability of the clusters thus can be analyzed individually. Focusing on still the $m$th cluster, transforming the variational equations into the mode space spanned by the submatrix $\mathbf{B}^m$, the $C_m-1$ modes can be further decoupled from each other, with the dynamics of each isolated mode described by the variational equation    
\begin{equation}
\delta \dot{\mathbf{z}}^m_{i'}=[D\mathbf{F}(\mathscr{X}_m)+\varepsilon \lambda^m_{i'} D\mathbf{H}(\mathscr{X}_m)]\delta \mathbf{z}^m_{i'},
\label{msf1}
\end{equation}
where $\delta \mathbf{z}^m_{i'}$ is the $i'$th mode and $0>\lambda_{1}^m\geq \lambda_{2}^m\geq\ldots \geq\lambda_{C_m-1}^m$ are the eigenvalues of the Laplacian matrix $\mathbf{B}^m-\mathbf{K}\mathbf{I}$. Here, $\mathbf{K}=\{k_i\}$, with $k_i=-\sum_{j'=1}^{C_m-1}b_{ij'}$ the coupling intensity of node $i$. $\mathbf{I}$ is the identity matrix. Noticing that nodes within the same cluster are synchronized to the same manifold, the stability of the transverse modes described by Eq. (\ref{msf1}) thus can be analyzed by the standard MSF approach \cite{MSF1,MSF2,MSF3}, as follows. Defining the generic coupling strength $\sigma\equiv -\varepsilon\lambda$, the MSF of $m$th cluster reads
\begin{equation}
\delta \dot{\mathbf{z}}^m=[D\mathbf{F}(\mathscr{X}_m)-\sigma D\mathbf{H}(\mathscr{X}_m)]\delta \mathbf{z}^m.
\label{msf}
\end{equation} 
 
\begin{figure}[tbp]
\includegraphics[width=0.8\textwidth]{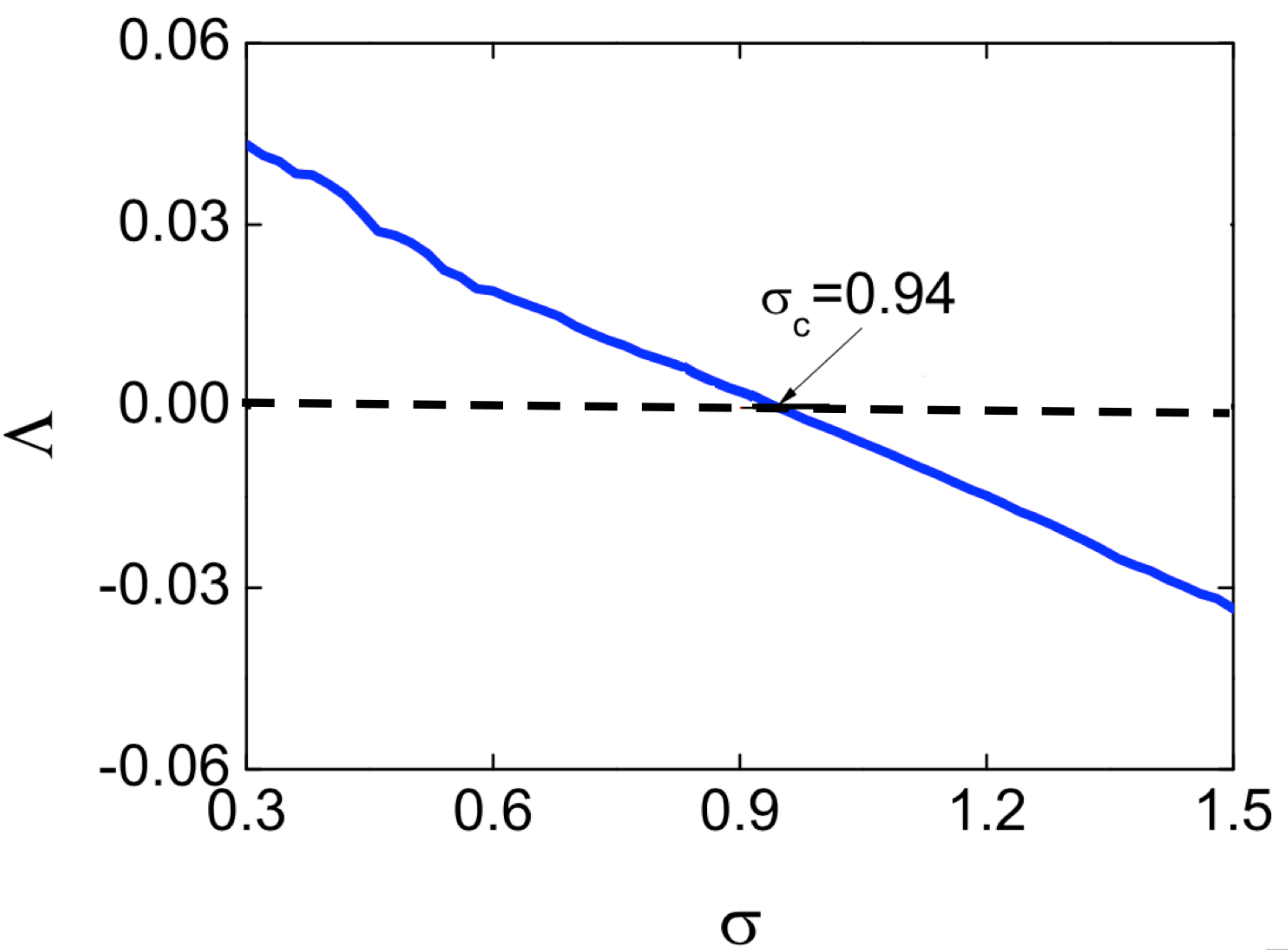}
\caption{(Color online) For the chaotic HR oscillators coupled through variable $x$, the variation of the largest Lyapunov exponent of the master stability function [Eq. (\ref{msf})], $\Lambda$, as a function of the generic coupling strength, $\sigma=\varepsilon\lambda$. $\Lambda$ is negative when $\sigma>\sigma_c\approx 0.94$.}
\label{fig4}
\end{figure}

Eq. (\ref{msf}) can be solved numerically, by which the dependence of the largest Lyapunov exponent $\Lambda$ as a function of $\sigma$ can be obtained, i.e., the MSF curve. For the chaotic neuron oscillator adopted in our studies, numerically it is found that $\Lambda<0$ when $\sigma>\sigma_c\approx 0.94$, as shown in Fig. \ref{fig4}. It should be noted that, dislike the situation of global synchronization where the synchronous manifold is governed by the same dynamics as the isolated oscillator [i.e. $\dot{\mathbf{X}}_s=\mathbf{F}(\mathbf{X}_s)$, with $\mathbf{X}_s$ the global synchronization manifold], in cluster synchronization the synchronous manifolds, $\mathscr{X}_m$, are determined by Eq. (6), which might have motions different from that of isolated oscillator. This situation occurs occasionally in systems of coupled oscillators (maps), where the motion of isolated oscillator is chaotic but becomes periodic when coupled together. However, for the chaotic HR oscillator employed in the present work, numerical evidences show that in the cluster synchronization state the synchronous manifolds, $\mathscr{X}_m$, have the same type of (chaotic) motion as the isolated oscillator. More specifically, although the synchronous manifolds are not synchronized, their statistical properties (i.e., the shape and location of the attractors, the fractal dimension, the distribution of the finite-time Lyapunov exponent, etc) can be approximately treated as identical to that of isolated oscillator. With this approximation, the stability of the clusters can be analyzed by the unified equation, i.e., Eq. (12). The stability condition for the $m$th cluster thus is $\sigma_1=-\varepsilon_m\lambda_1^m>\sigma_c$, which gives $\varepsilon_m>-\sigma_c/\lambda_1^m$. Similarly, we can obtain the stability conditions for the other clusters, $\varepsilon_{m'}>-\sigma_c/\lambda_1^{m'}$. Let $\lambda^B_{max}$ be the largest eigenvalue of $\mathbf{B}$, then the necessary condition for all the synchronous clusters to be stable is
\begin{equation}
\varepsilon>\varepsilon_b=-\sigma_c/\lambda_{max}^B.
\label{cond1}
\end{equation} 
In the meantime, to avoid the trivial situation of global synchronization, we need to keep at least two modes of the synchronous space unstable. That is, besides the mode of the synchronous manifold (the mode associated with the null eigenvalue of the coupling matrix), we should have $\Lambda>0$ for at least one other mode in the space of $\mathbf{D}$. Let $0=\lambda^D_1>\lambda^D_2\geq \lambda^D_3\geq \ldots \geq \lambda^D_M$ be the eigenvalues of the Laplacian matrix $\mathbf{D}-\mathbf{K}\mathbf{I}$ [$\mathbf{K}=\{k_i\}$, with $k_i=-\sum_{j'=1}^M d_{ij'}$, and $\mathbf{I}$ the identity matrix, to fulfill the above requirement, we should have $\Lambda(-\varepsilon \lambda_2^D)>0$. As shown in Fig. \ref{fig4}, $\Lambda$ is negative for $\sigma>\sigma_c$, we therefore have the $2$nd criterion for generating stable cluster synchronization
\begin{equation}
\varepsilon<\varepsilon_d=-\sigma_c/ \lambda_2^D.
\label{cond2}
\end{equation}
The  two conditions, Eqs. (\ref{cond1}) and (\ref{cond2}), can be physically interpreted, as follows. Eq. (\ref{cond1}) guarantees that the coupling strength is larger enough to confine the oscillators in each cluster onto the same trajectory, while Eq. (\ref{cond2}) ensures that the coupling strength is weaker enough for not generating global synchronization. To generate stable cluster synchronization, both conditions should be satisfied. [The critical coupling strength for global synchronization can be obtained from the MSF method \cite{MSF1,MSF2,MSF3}, $\varepsilon_c=\sigma_c/\lambda_{max}$, with $\lambda_{max}=\max\{\lambda_{max}^B,\lambda_2^D\}$ (i.e., the largest non-zero eigenvalue of the matrix $\mathbf{A}-\mathbf{K}\mathbf{I}$).]

\subsection{Theoretical analysis of the experimental results}

By the above criteria, we now are able to analyze the synchronization behaviors observed in experiments (i.e., the results presented in Figs. \ref{fig2} and \ref{fig3}). For the original network presented in Fig. \ref{fig1}(a), the network structure satisfies the reflection symmetry $\mathbf{S}_1$. This symmetry supports potentially the synchronous pattern $(a,b,c,c,b)$, in which each symbol represents a unique trajectory and the symbols are ordered according to the node index. For instance, nodes $2$ and $5$ are of identical trajectory, as they are represented by the same symbol $b$. According to symmetry $\mathbf{S}_1$, we can construct the following permutation matrix

\begin{equation}
\mathbf{R}_{S1}=\left(
  \begin{array}{cccccc}
    1 & 0 & 0 & 0 & 0\\
    0 & 0 & 0 & 0 & 1 \\
    0 & 0 & 0 & 1 & 0 \\
    0 & 0 & 1 & 0 & 0 \\
    0 & 1 & 0 & 0 & 0 \\
    \end{array}
\right).
\label{pm}
\end{equation}
By the eigenvectors of $\mathbf{R}_{S1}$, we have the transformation matrix
\begin{equation}
\mathbf{T}_s=\frac{1}{\sqrt{2}}\left(
  \begin{array}{cccccc}
         0      &  0  &     \sqrt{2}     &  0  &      0    \\
        0  &    1      &      0     &      1     &  0 \\
         -1      &     0      &      0     &      0     &      1    \\
      1  &     0      &      0     &      0     &  1 \\
         0      &  -1 &      0     &   1 &      0     \\
   \end{array}
\right).
\label{tm}
\end{equation}
By the transformation $\mathbf{A}'=\mathbf{T}_s^{-1}A\mathbf{T}_s$, we get the submatrices
\begin{equation}
\mathbf{B}=\left(
  \begin{array}{cc}
      -4 & -1 \\
       -1 & -4 \\
  \end{array}
\right)
\end{equation}
and
\begin{equation}
\mathbf{D}=\left(
  \begin{array}{ccc}
      -4 & \sqrt{2}  & \sqrt{2} \\
      \sqrt{2}  & -2 & 1        \\
      \sqrt{2} & 1 & -2  \\
  \end{array}
\right).
\end{equation}
The largest eigenvalue of $\mathbf{B}$ is $\lambda_{max}^B=-3$, which, according to Eq. (\ref{cond1}), gives the $1$st condition for generating the pattern $(a,b,c,c,b)$, $\varepsilon>\varepsilon_b\approx 0.31$. For the submatrix $\mathbf{D}$, we have $\lambda_2^D=-3$, which, according to Eq. (\ref{cond2}), requires that $\varepsilon<\varepsilon_d\approx 0.31$. Since $\varepsilon^{exp}\approx 0.35>\varepsilon_d$, the $2$nd condition thus is not satisfied, indicating that the synchronous pattern $(a,b,c,c,b)$ is unstable, as depicted in Fig. \ref{fig2}(b). (In fact, because of $\varepsilon_d=\varepsilon_b$, it can be predicated theoretically that cluster and global synchronization occur simultaneously at the point $\varepsilon_c\approx 0.31$.  This predication is in agreement with the experimental observation that the network reaches global synchronization when $\varepsilon^{exp}>\varepsilon^{exp}_c\approx 0.33$.)

When the shortcut link between nodes $2$ and $5$ is removed [Fig. \ref{fig2}(c)], the reflection symmetry $\mathbf{S}_1$ is not affected [which supports still the synchronous pattern $(a,b,c,c,b)$]. As such, the permutation matrix is identical to Eq. (\ref{pm}), so is the transformation matrix $\mathbf{T}_s$ [Eq. (\ref{tm})]. Transforming the coupling matrix $\mathbf{A}$ of the network shown in Fig. \ref{fig2}(c) into the mode space, we have the submatrices 
\begin{equation}
\mathbf{B}=\left(
  \begin{array}{cc}
      -4 & -1 \\
       -1 & -2 \\
  \end{array}
\right)
\end{equation}
and
\begin{equation}
\mathbf{D}=\left(
  \begin{array}{ccc}
      -4 & \sqrt{2}  & \sqrt{2} \\
      \sqrt{2}  & -2 & 1        \\
      \sqrt{2} & 1 & -2  \\
  \end{array}
\right).
\end{equation}
For $\mathbf{B}$, we have $\lambda_{max}^B\approx -1.6$, which gives $\varepsilon_b\approx 0.59$; while for $\mathbf{D}$, we have $\lambda_2^D=-3$, which indicates that $\varepsilon_d\approx 0.31$. In experiment, we have $\varepsilon^{exp}\approx 0.35$ [Fig. \ref{fig2}(d)]. Since $\varepsilon^{exp}<\varepsilon_b$, the first criterion for cluster synchronization, i.e., Eq. (\ref{cond1}), therefore is not satisfied. As a result of this, the synchronous pattern $(a,b,c,c,b)$ is not observable. Meanwhile, as $\varepsilon^{exp}<\varepsilon_c\approx 0.59$, the network is neither reaching global synchronization. This explains the completely desynchronized state observed in experiment [Fig. \ref{fig2}(d)].

We finally analyze the stability of the cluster synchronization state in the network shown in Fig. \ref{fig3}(a), i.e., when the shortcut link between nodes $1$ and $3$ is removed. For this network, its structure satisfies the reflection symmetry $\mathbf{S}_2$, which is different from the reflection symmetry $\mathbf{S}_1$. The corresponding permutation matrix reads  
\begin{equation}
\mathbf{R}_{S2}=\left(
  \begin{array}{cccccc}
    0 & 0 & 0 & 0 & 1\\
    0 & 0 & 0 & 1 & 0 \\
    0 & 0 & 1 & 0 & 0 \\
    0 & 1 & 0 & 0 & 0 \\
    1 & 0 & 0 & 0 & 0 \\
    \end{array}
\right),
\label{pm}
\end{equation}
and the transformation matrix reads
\begin{equation}
\mathbf{T}_s=\frac{1}{\sqrt{2}}\left(
  \begin{array}{cccccc}
         0      &  1  &      0     &  1  &      0    \\
    -1   &     0      &      0     &      0     &  1 \\
         0      &     0      &      1     &      0     &      0    \\
      1  &     0      &      0     &      0     &  1 \\
         0      &  -1 &      0     &   1 &      0     \\
   \end{array}
\right).
\label{tm}
\end{equation}
By the transformation $\mathbf{A}'=\mathbf{T}_s^{-1}A\mathbf{T}_s$, we have the submatrices
\begin{equation}
\mathbf{B}=\left(
  \begin{array}{cc}
      -3 & 0 \\
       0 & -4 \\
  \end{array}
\right)
\end{equation}
and
\begin{equation}
\mathbf{D}=\left(
  \begin{array}{ccc}
      -2 & 0  & \sqrt{2} \\
      0  & -2 & 2        \\
      \sqrt{2} & 2 & -3  \\
  \end{array}
\right).
\end{equation}
The largest eigenvalue of $\mathbf{B}$ is $\lambda_{max}^B=-3$, which, gives $\varepsilon_b\approx 0.31$. For the submatrix $\mathbf{D}$, we $\lambda_2^D=-2$, which gives $\varepsilon_d\approx 0.47$. As $\varepsilon_b<\varepsilon^{exp}\approx 0.35<\varepsilon_d$, both conditions for stable cluster synchronization are satisfied [Eqs. (\ref{cond1}) and (\ref{cond2})]. The synchronous pattern $(a,b,c,b,a)$ thus is judged as stable, as confirmed by the experiment [Figs. \ref{fig3}(b-d)]. 

The roles of shortcut links on cluster synchronization now can be interpreted, as follows. If the removal of the shortcut link does not affect the network symmetry, the distribution of the synchronous clusters, namely the synchronous pattern, is not changed, yet the stability of this pattern will be adjusted. This is the case when the network structure is changed from Fig. \ref{fig2}(a) to Fig. \ref{fig2}(c), where the removal of the link between nodes $2$ and $5$ does not change the reflection symmetry $\mathbf{S}_1$, but the network dynamics is changed from global synchronization to complete desynchronization. On the other hand, the removal of of the shortcut link may alter the network symmetry. This is what happens when the link between nodes $1$ and $3$ is removed [Fig. \ref{fig3}(a)], where the modified network satisfies the different reflection symmetry $\mathbf{S}_2$. It is for just this new symmetry $\mathbf{S}_2$ that stable cluster synchronization can be generated. 

\section{DISCUSSIONS AND CONCLUSION}

As discussed in Sec. \ref{theory}, the emergence of cluster synchronization relies strictly on the network symmetry. For the sake of simplicity, in theoretical analysis we have assumed that the network is constituted by identical oscillators and by unweighted links. In realistic systems, mismatches in the parameters of the oscillators and in the weights of the network links are unavoidable, resulting in non-perfect network symmetries. This raises the general concern that cluster synchronization may not be generated in realistic systems, especially for networks of complicated nodal dynamics \cite{SynPatPecora}. In addition, the generation of cluster synchronization requires not only the synchronization of each cluster [Eq. (\ref{cond1})], but also the desynchronization among the clusters [Eq. (\ref{cond2})]. As such, comparing to global synchronization, cluster synchronization is normally believed as more difficult to be generated \cite{CSPecora2014}. Our experimental study shows that, despite of the above concerns (non-perfect topological symmetry, noise perturbations, and chaotic nodal dynamics), robust cluster synchronization can still be generated in experimental complex networks. The current study provides new evidence on the generation of cluster synchronization in realistic situations, and gives confidence for finding cluster synchronization in large-size complex networks, e.g., the human brain system.

The theoretical framework we have proposed for analyzing cluster synchronization is general, and is independent of the details of the network models. For instance, by changing the nodal dynamics or the coupling function, the MSF curve may have bounded stable region \cite{MSF2,MSF3}. According to the theoretical framework, it is straightforward to find that in this case the stability of cluster synchronization requires one additional criterion, $\Lambda(-\varepsilon\lambda_{min}^B)<0$, with $\lambda_{min}^B$ the smallest eigenvalue calculated from the submatrix $\mathbf{B}$. More importantly, this framework may potentially be applied to complex networks consisting of non-identical oscillators. For the sake of simplicity, we have assume in our model that all oscillators in the network are of the same type of dynamics. Yet, as implied by Eqs. (\ref{eq9a}) and (\ref{eq9b}), the synchronous manifolds of the clusters are not necessary to be following the same type of dynamics, as in the mode space spanned by the permutation matrix the clusters are completely decoupled from each other. This gives indications on how to realize cluster synchronization in complex network constituted by different types of oscillators: keeping the same type of oscillators symmetric and arranging them inside the same cluster, an interesting phenomenon warranting further investigation. 

In summary, we have investigated, experimentally and theoretically, the synchronization behavior of a small-size complex network consisting of chaotic electrical circuits. By setting the network at the globally synchronizable state initially \cite{CSOTT2007}, we have adjusted the network structure slightly be removing one of the shortcut links and found that the network might either be completely desynchronized or be partially desynchronized. When the network is partially desynchronized, it is found that the oscillators are synchronized into different groups, showing the phenomenon of cluster synchronization. We have conducted a detailed analysis on the stability of the cluster synchronization state, and derived explicitly the criteria for cluster synchronization. The theoretical analysis well explains the experimental results, and reveals in depth the roles of the shortcut links in affecting the formation of synchronous patterns. Our study sheds new lights on the interplay between network dynamics and structure, and is a step forward to the full exploration of synchronous patterns in the large-size, realistic networks.  

This work was supported by the National Natural Science Foundation of China under the Grant No.~11375109, and by the Fundamental Research Funds for the Central Universities under the Grant No.~GK201601001. YZY and XGW thank the support from the National Demonstration Center for Experimental x-physics education (Shaanxi Normal University).

\end{document}